\documentclass[11pt]{JHEP3}

\newcommand{\lbl}[1]{\label{#1}}

\usepackage{bm}

\usepackage{amsmath}
\usepackage{graphicx}
\usepackage{amssymb}
\usepackage{mathrsfs}
\usepackage{bbm}

\newcommand{\larr}[1]{\xrightarrow{\!\!{\scriptscriptstyle{#1}}\!\!} }
\newcommand{\be}{\begin{eqnarray}}
\newcommand{\ee}{\end{eqnarray}}
\newcommand{\eq}[1]{(\ref{#1})}

\newcommand{\inter}{{\mathrm{int}}}
\newcommand{\WS}{{{\mathrm{WS}}}}

\newcommand{\cY}{{\mathcal{Y}}}

\newcommand{\cL}{{\cal L}}

\newcommand{\bbbone}{{{\mathchoice {\rm 1\mskip-4mu l} {\rm 1\mskip-4mu l}
{\rm 1\mskip-4.5mu l} {\rm 1\mskip-5mu l}}}}

\newcommand{\vct}[1]{\vec{#1}}
\newcommand{\vecW}{\vct{W}}
\newcommand{\vecG}{\vct{G}}

\title{Non-Abelian Supercurrents And de Sitter Ground State In Electroweak Theory}

\author{$^{a,b}$M.N. Chernodub,
$^c$Ludvig Faddeev and  $^{a,d,e}$Antti J. Niemi  \\
$^a$Laboratoire de Mathematiques et Physique Theorique
CNRS UMR 6083, F\'ed\'eration Denis Poisson, Universit\'e de Tours,
Parc de Grandmont, F37200, Tours, France \\
$^b$ITEP, B. Cheremushkinskaya 25, Moscow 117218, Russia \\
$^c$S. Petersburg Branch of Steklov Mathematical Institute,
Russian Academy of Sciences, Fontanka 27, St. Petersburg, Russia\\
$^d$Department of Physics and Astronomy, Uppsala University,
P.O. Box 803, S-75108, Uppsala, Sweden \\
$^e$Chern Institute of Mathematics, Tianjin 300071, P.R. China \\
E-Mail: {\tt Maxim.Chernodub@lmpt.univ-tours.fr, Faddeev@pdmi.ras.ru,
Antti.Niemi@physics.uu.se}}

\abstract{We show that gauge symmetry breaking in the Weinberg-Salam model
can be implemented by a mere change of variables and without any explicit gauge
fixing. The change of variables entails the concept of
supercurrent which has been widely employed in the study of
superconductivity. It also introduces a separation
between the isospin and the hypercharge, suggesting
that our new variables describe a strongly coupled
regime of the electroweak theory. We discuss the description
of various embedded topological defects in terms of these variables.
We also propose that in terms of our variables
the Weinberg-Salam model can be interpreted in terms of a
gravity theory with the modulus of Higgs field as dilaton and
the de Sitter space as the ground state.}

\preprint{UUITP-04/08, ITEP-LAT/2008-10}


\begin{document}

\section{Introduction}

Experiments with the LHC accelerator at CERN
may soon reveal the mechanism of the
electroweak symmetry breaking, pivotal to our
understanding of fundamental interactions. The symmetry breaking
is supposed to proceed in a textbook manner: The Higgs
field develops a constant ground state expectation value which
gives a mass for the charged $W^\pm$ and the neutral $Z$
bosons but leaves the photon massless.

However, there remains theoretical issues to be addressed.
Among them, a theorem by
Elitzur~\cite{ref:Elitzur} states that
it should be impossible to spontaneously break a local symmetry
such as the electroweak gauge group
$G_\WS = SU_L(2)\times U_Y(1)$. According
to this theorem only global symmetries can be spontaneously
broken. A gauge fixed theory avoids this conundrum
since the local gauge symmetry becomes explicitly broken by
the gauge fixing condition.
Furthermore, both numerical lattice
simulations~\cite{ref:lattice:phase}
and formal arguments~\cite{ref:lattice:general} show that
the transition between the symmetric and the Higgs phase can proceed
in an analytic manner along a continuous path in the phase
diagram. In particular, since the gauge symmetry is unbroken in
the symmetric phase it must remain unbroken also in the Higgs phase,
and this appears to bring the Higgs mechanism of the Weinberg-Salam
model in line with Elitzur's theorem.

Here we address these issues from a new perspective by showing how
the entire electroweak Lagrangian can be written
in terms of manifestly $ SU_L(2)\times U_Y(1)$
invariant variables that are the
analogs of the Mei\ss ner supercurrent in the Ginzburg-Landau approach
to BCS superconductor. These variables can be interpreted in
terms of spin-charge separation, in line with the spin-charge
separation that has been previously employed in the context of
strongly correlated electron systems in condensed matter
physics \cite{anderson}, \cite{walet}, \cite{faddeco}.
Thus the proper interpretation of our variables appears to be in terms
of the strongly coupled (strongly correlated) dynamics of electroweak theory.
Furthermore, since the non-Abelian supercurrents implement the effects
of gauge symmetry breaking by a mere change of variables and
without any gauge fixing,
any issues with Elitzur's theorem become obsolete.

We also show that the isospin-hypercharge separated electroweak
Lagrangian can be given a gravitational interpretation in terms
of conformal geometry. This suggests a dual relation between the
strongly coupled electroweak theory and a theory of gravitation
\cite{polch}. In particular, we propose that the ground state
of the electroweak theory is the four dimensional
de Sitter space, with the modulus of the Higgs field as the dilaton.

\section{Abelian Higgs Model}

We start by illustrating our proposal by considering a complex
scalar field $\phi$ and a vector field  $A_i$ in three space
dimensions, in the context of the conventional Landau-Ginzburg approach to
BCS  superconductivity. There are a total of five
independent fields. We
introduce an invertible change of variables to a
set of five independent fields $(J_i, \rho, \theta)$,
\begin{equation}
\begin{array}{lcl}
\phi \  & \to & \ \rho \cdot e^ {i \theta } \\
A_i \ & \to & J_i = \frac{i}{4 e
|\phi|^2}\left[ \phi^* ( \partial_i - 2 i e A_i ) \phi - c.c. \right]
\end{array}
\lbl{asu}
\end{equation}
Note that we have not yet detailed any physical model
where these variables appear as field degrees of freedom.
We now proceed to the Landau-Ginzburg Hamiltonian which is relevant to
BCS superconductivity, with $\phi$ the scalar field that describes
Cooper pairing of electrons and $A_i$ the (Maxwellian) $U(1)$
magnetic vector potential,
\begin{equation}
{\mathcal H} = \frac{1}{2} B^2_{i} + |(\partial_i - 2i e A_i)
\phi|^2 + \lambda \left(|\phi|^2 -
v^2
\right)^2\,.
\lbl{H1}
\end{equation}
Here $B_i$ denotes the magnetic field.
This Hamiltonian displays the familiar Maxwellian $U(1)$
gauge invariance
\[
\begin{matrix}
\phi \ \to \ e^{2ie\eta} \phi  \\
A_i \ \to \ A_i + \partial_i \eta \end{matrix}
\]
In terms of the new fields (\ref{asu}) the Hamiltonian (\ref{H1}) is
\begin{equation}
{\mathcal H} = \frac{1}{4} \left(
J_{ij} + \frac{\pi}{e} \widetilde\sigma_{ij}
\right)^2
\!\!
+ (\partial_i \rho)^2 + \rho^2 J_i^2
+ \lambda \left(\rho^2 - \eta^2 \right)^2\!\!,\!\!
\lbl{H2}
\end{equation}
with
\[
J_{ij} = \partial_i J_j - \partial_j J_i
\]
and
\be
\widetilde\sigma_{ij} = \epsilon_{ijk} \sigma_k =
\frac{1}{2\pi}\, [\partial_i , \partial_j ] \theta
\lbl{ds1}
\ee
Here $\sigma_i$ is the string current, its support in $\mathbb R^3$
coincides with the worldsheet
of the core of a
nonrelativistic Abrikosov vortex.
When (\ref{H2}) describes such a vortex, (\ref{ds1}) subtracts a singular
contribution that emerges from $J_{ij}$. This singularity also
emanates in the third term in the {\it r.h.s.} of (\ref{H2}). There it
becomes removed by the density $\rho$ which
vanishes on the worldsheet of the vortex core.

The Hamiltonian (\ref{H2}) involves only variables
that are manifestly $U(1)$ gauge invariant, there is no local
gauge invariance present in (\ref{H2}). But no gauge has been fixed
in deriving (\ref{H2}) from (\ref{H1}). Instead, all
gauge dependent quantities have been explicitly
eliminated by the change of variables.
Notice that Eq.~(\ref{ds1}) is invariant under a $U(1)$
gauge transformation that
entails a shift in $\theta$ by a twice differentiable
scalar function.
Since (\ref{ds1}) displays no gauge invariances,
there are no issues with Elitzur's theorem.
Moreover, since $\rho \geq 0$ there are no gauge invariant
global symmetries to be spontaneously broken by the potential term
even though the Mei\ss ner effect does reflect the properties
of the potential term.

\section{Non-Abelian Supercurrents}

We wish to generalize the previous approach to the (bosonic sector
of the) standard electroweak theory, defined by the classical
Lagrangian
\begin{equation}
\cL_\WS
= \frac{1}{4} \vec G_{\mu\nu}^2(W) + \frac{1}{4} F^2_{\mu\nu}(Y)
+ | D_\mu { \Phi}|^2 + \lambda |\Phi|^4 + \mu^2 |\Phi|^2
\lbl{ew:lag}
\end{equation}
We use the notation of \cite{abers}. For the moment
we work in a spacetime with Euclidean signature. The
matrix-valued $SU_L(2)$ isospin gauge field is
\[
\widehat W_\mu \equiv  W^a_\mu \tau^a = {\vec {W}}_\mu \cdot \vec \tau
\]
with $\tau^a$ the isospin Pauli matrices, $Y_\mu$ is the (Abelian)
$U_Y(1)$ hypergauge field, and
\be
\vecG_{\mu\nu}(W) & = & \partial_\mu \vecW_\nu -
\partial_\nu \vecW_\mu - g \, \vecW_\mu \times \vecW_\nu\,,
\lbl{eq:G}\\
F_{\mu\nu}(Y) & = & \partial_\mu Y_\nu - \partial_\nu Y_\mu\, .
\lbl{eq:F}
\ee
The $SU_L(2) \times U_Y(1)$ covariant derivative is
\begin{equation}
D_\mu = \bbbone \, \partial_\mu - i \frac{g}{2}
\widehat W_\mu  - i \frac{g'}{2} Y_\mu\,\bbbone\,,
\lbl{covdev}
\end{equation}
where $\bbbone$ is the $2 \times 2$ unit matrix in the isospin space.
The complex isospinor Higgs field $\Phi$ is decomposed as follows,
\begin{equation}
\Phi \ = \phi \, \mathcal X \ \ \ \ {\rm with}
\quad \ \ \
\phi = \rho \, e^{i\theta} \
\ \ \ \& \ \ \ \mathcal X = \
\mathcal U
\left( \begin{matrix} 0 \\ 1 \end{matrix} \right).
\lbl{phi}
\end{equation}
Here $\phi$ is a complex field,  $\mathcal X$
a two-component complex isospinor, and $\mathcal U $ a
$SU_L(2)$ matrix. The
\[
G_{WS} = SU_L(2) \times U_Y(1)
\]
gauge transformation
acts on $\Phi$ as follows,
\begin{equation}
\Phi \larr{G_\WS}e^{i \omega_Y} \Omega \, \Phi
\ \  \Rightarrow \ \ \left\{ \begin{matrix}
\phi \hskip 2mm \ \longrightarrow e^{i \omega_Y} \phi
\\
\mathcal X \longrightarrow \ \Omega \mathcal X \end{matrix} \right.
\lbl{eq:splitting}
\end{equation}
where $\Omega \in SU_L(2)$ and $e^{i\omega_Y} \in U_Y(1)$. As a
consequence the decomposition separates isospin from hypercharge
\cite{ref:grav}.
It also introduces a new (internal) {\it compact} gauge group
\begin{equation}
U_\inter(1):\quad \begin{matrix}
\phi \to e^{- i \omega_c} \phi \\
\mathcal X \to e^{i \omega_c} \mathcal X \end{matrix}
\lbl{eq:internal}
\end{equation}
which leaves the field $\Phi$ intact.
The spinor $\mathcal X \equiv \mathcal X_1$ and its
isospin conjugate
\[
\mathcal X_2 = e^{i\beta}
i \tau_2 { {\mathcal X}}_{}^*
\]
form an orthonormal basis
($i,j=1,2$ and $a,b = \uparrow, \downarrow$),
\[
\mathcal X_i^\dagger \cdot \mathcal X_j \
\equiv \sum_{a = \uparrow, \downarrow} \mathcal X_{ia}^* \mathcal X_{aj}
\ = \ \delta_{ij}
\]
\[
\sum_{i=1,2}
\mathcal X_{i a}^{} \mathcal X_{ib}^\dagger \ = \ \delta_{ab}
\]
Hereafter we set $\beta=0$ as it parameterizes an internal
degree of freedom that was already accounted for by~\eq{eq:internal}.

We introduce the non-Abelian supercurrents in parallel with Eq.~\eq{asu}.
For this we expand the covariant derivative of the Higgs field
in the spinor basis ($\mathcal X_1,\mathcal X_2$),
\begin{equation}
D_\mu \Phi \ = \ \Bigl[\frac{1}{\rho}\partial_\mu \rho + \frac{i}{2}
\Bigl( g J_{\mu}^3 - g' \mathcal Y_\mu^{} \Bigr) \Bigr] \, \Phi
- i \frac{g}{2} J^+_\mu \cdot \Phi_c
\lbl{DPhi}
\end{equation}
Here
\[
\Phi_c = \phi \mathcal X_2
\]
is the isocharge conjugated Higgs field.
The supercurrents
$J^+_\mu$, $J_{\mu}^3$ and $\mathcal Y_\mu$
emerge when we project out the spinor components
\begin{eqnarray}
J_\mu^+ \! & = & \! \frac{2i}{g}\mathcal X_{2}^\dagger \!
\left( \partial_\mu - \frac{ig}{2} \widehat W_\mu \right) \! \mathcal X_1^{}
\equiv
\vec W_\mu \cdot \vec e_+ + \frac{i}{g} \vec e_3 \cdot \partial_\mu \vec e_+\,,
\lbl{J+}\qquad \\
J^3_\mu \! & = & \! \frac{2}{i g}
\mathcal X_1^\dagger
\! \left( \partial_\mu - \frac{ig}{2} \widehat W_\mu \right) \!
\mathcal X_1^{}
\equiv
\vec W_\mu \cdot \vec e_3 - \frac{i}{2g} \vec e_- \cdot
\partial_\mu \vec e_+ \, ,
\lbl{J3}\qquad \\
\mathcal Y_\mu \!\! & = & \! \frac{i}{g' |\phi|^2}
\Bigl[ \phi^\star \Bigl(\partial_\mu - i
\frac{g'}{2} Y_\mu \Bigr)\phi - c.c. \Bigr] \,.
\lbl{Y}
\end{eqnarray}
Here
\[
J^+_\mu = J^1_\mu + i J^2_\mu
\]
and
\[
\vec e_+ \equiv
\vec e_1 + i \vec e_2
\]
with $\vec e_i  \ (i=1,2,3)$ three mutually
orthogonal unit vectors,
\begin{eqnarray}
\vec e_3 & = & - \frac{\Phi^\dagger {\hat{\vec\tau}} \Phi}{\Phi^\dagger \Phi}
         \equiv - \mathcal X_1^\dagger {\hat{\vec\tau}} \mathcal X_1 \\
\vec e_+ & = & \mathcal X_2^\dagger {\hat{\vec\tau}} \mathcal
X_1
\end{eqnarray}
and the internal gauge group (\ref{eq:internal})
acts as follows,
\begin{equation}
U_\inter(1): \quad \begin{matrix} \vec e_3 \to \vec e_3 \\
\quad \vec e_+ \to e^{2 i \omega_c} \vec e_+ \end{matrix}
\lbl{eq:internal:e}
\end{equation}

In parallel with the Abelian Higgs model we
interpret (\ref{J+})-(\ref{Y}) as a change
of variables from the original twenty (real) fields
$(W_\mu^a, Y_\mu, \Phi)$ to a new set of twenty fields.
In addition of the sixteen
$(J^a_\mu, \mathcal Y_\mu)$ these include the
modulus $\rho$ and the orthogonal triplet $\vec e_i$.

The supercurrents $(J^3_\mu , J^+_\mu , \mathcal Y_\mu)$
are the manifestly $G_\WS = SU_L(2) \times U_Y(1)$ gauge
invariant {\it electroweak supercurrents}. They are the
non-Abelian generalizations of~\eq{asu}. But
under the internal gauge symmetry~\eq{eq:internal}
supercurrents $(J^3_\mu , J^+_\mu , \mathcal Y_\mu)$
transform nontrivially, in the following manner:
\begin{equation}
U_\inter(1):
\qquad
\begin{array}{rcl}
J^+_\mu & \to & e^{2 i \omega_c} J^+_\mu \,, \\[1mm]
J^3_\mu & \to & J^3_\mu + \frac{2}{g} \partial_\mu \omega_c \,,\\[1mm]
\cY_\mu & \to & \cY_\mu + \frac{2}{g'} \partial_\mu \omega_c \,.
\end{array}
\label{eq:internal:J}
\end{equation}

Finally, we note that quite similar electroweak supercurrents have been
previously presented in \cite{shapo}. See also \cite{volovik} for
a related construction.

\section{ Electroweak Lagrangian in Supercurrent
Variables }

In terms of the $G_{WS} = SU_L(2)\times U_Y(1)$ invariant
variables the classical electroweak Lagrangian~(\ref{ew:lag})
acquires a form  similar to (\ref{H2}),
\begin{eqnarray}
& &
\hskip -4mm
\cL_\WS =
\frac{1}{4} \left( \! \vec G_{\mu\nu}(\vec J)
+ \frac{4 \pi}{g} {\vec{\widetilde \Sigma}}_{\mu\nu} \!
\right)^2 \! \! \! + \frac{1}{4} \left( \!
F_{\mu\nu}(\mathcal Y)
+ \frac{4\pi}{g'} \widetilde \sigma_{\mu\nu}^\phi \!
\right)^2
\nonumber\\
& & \hskip -5mm
+ (\partial_\mu \rho)^2 \!
+ \frac{\rho^2}{4}\left( g J^3_\mu - g' \mathcal Y_\mu\right)^2
\! + \frac{\rho^2 g^2}{4} J_\mu^+ J_\mu^-
+ \lambda \hskip 0.6mm \rho^4 + \mu^2 \rho^2
\lbl{wsgi}
\end{eqnarray}
Here $\vec G_{\mu\nu}$ and $F_{\mu\nu}$ are
now the curvatures of $\vec J_\mu$ {\it resp.} $\mathcal Y_\mu$,
\be
\vecG_{\mu\nu}(\vec J) & = & \partial_\mu \vec J_\nu -
\partial_\nu \vec J_\mu - g \, \vec J_\mu \times \vec J_\nu\,,
\lbl{eq:GJ}\\
F_{\mu\nu}(\mathcal Y) & = & \partial_\mu \mathcal
Y_\nu - \partial_\nu \mathcal Y_\mu\, .
\lbl{eq:FY}
\ee
The $G_\WS$--invariant dual string tensor
\begin{equation}
\widetilde \sigma_{\mu\nu}^\phi = \frac{1}{2
\pi} \,[\partial_\mu , \partial_\nu ] \, \arg \phi
\lbl{ds2}
\end{equation}
describes the embedding of (singular)
stringlike vortex cores in (\ref{wsgi}) in analogy with
(\ref{ds1}). Its non-Abelian and $G_\WS$--invariant extension
generalizes Eq.~\eq{ds2} to $SU_L(2)$,
\begin{equation}
{\widetilde \Sigma}_{\mu\nu}^i \!\! =  \!\!
\frac{i}{\pi} {\mathrm{Tr}}\,
\Bigl[{\hat \tau}^i \Bigl({\mathcal U}^\dagger
[\partial_\mu,\partial_\mu] {\mathcal U}\Bigr)\Bigr]
\equiv
- \frac{1}{8\pi} \epsilon^{ijk} \bigl({\vec e}^{\hskip 0.3mm j}
\cdot [\partial_\mu, \partial_\nu] \, {\vec e}^{\hskip 0.5mm k} \bigr)
\lbl{calG}
\end{equation}
The (singular) codimension two surfaces described by (\ref{calG})
in $\mathbb R^4$ are world lines of stringlike vortex cores.
The boundaries of these surfaces
are curves in $\mathbb R^4$
that describe the world lines of pointlike
structures including the cores of Wu-Yang type
magnetic monopoles.

Note that in (\ref{wsgi})
the orthogonal triplet $\vec e_i$ only appears thru
the topological structures that are described
by the tensors (\ref{ds2}) and (\ref{calG}).
This leaves us with seventeen regular and manifestly
$SU_L(2) \times U_Y(1)$ gauge invariant field variables:
The original gauge symmetry has been
entirely eliminated by the change of variables and
without any gauge fixing.
The only surviving local gauge invariance
of (\ref{wsgi}) is the novel internal $U_\inter(1)$ gauge
symmetry (\ref{eq:internal:J}), that now defines
the Maxwellian gauge symmetry of (\ref{wsgi}).
In particular, since there are no local remaining
gauge symmetries to be broken there are no issues with Elitzur's theorem.
Furthermore, since $\rho \geq 0$ there are no discrete
symmetries to be broken by the {\it v.e.v.} of $\rho$. But if the
minimum of the potential occurs at $\rho = 0$ the supercurrents
are massless, and if the minimum occurs at $\rho \not =0$
three of the supercurrents acquire a nonvanishing mass.

Note that if one overlooks topological structures the functional
form of the Lagrangian
(\ref{wsgi}) coincides with that of the
original Lagrangian in the unitary gauge, even though here
no gauge fixing has taken place.

The $G_\WS$--gauge invariant $W$--bosons are
$W^\pm_\mu = J^\pm_\mu$, and $Z$--bo\-son and photon $A_\mu$ are
\begin{eqnarray}
Z_\mu & = & \cos\theta_W\, J^3_\mu - \sin\theta_W\, \mathcal Y_\mu \,, \\
A_\mu & = & \sin\theta_W\, J^3_\mu + \cos\theta_W\, \mathcal Y_\mu\,,
\lbl{eq:ZA}
\end{eqnarray}
where $\theta_W$ is the Weinberg angle,
\[
\sin\theta_W = \frac{g'}{\sqrt{ g^2 + {g'}^2} }
\]
Under the internal  $U_\inter(1)$
gauge symmetry~\eq{eq:internal:J} the $W$--boson
field transforms as a charged vector field.
From (\ref{eq:internal:J}) and (\ref{eq:ZA}) we conclude
that the $Z$-boson is a singlet while for the photon
we get
\begin{equation}
U_\inter(1):
\qquad
A_\mu  \to A_\mu + \frac{2}{g' g} \sqrt{g^2 +
g'^2} \, \partial_\mu \, \omega_c\,.
\lbl{eq:internal:A}
\end{equation}
In the case of a singular $\omega_c$ this gauge transformation
acts on the Abelian field strength tensor as follows,
\begin{equation}
F_{\mu\nu} \equiv \ \partial_{[\mu,} A_{\nu]} \to F_{\mu\nu} +
\frac{4\pi }{g' g} \cdot
n  \sqrt{g^2 + g'^2} \, {\widetilde \sigma}^{\mathrm{Dirac}}_{\mu\nu}\,,
\lbl{eq:f:law}
\end{equation}
where
\[
{\widetilde \sigma}^{\mathrm{Dirac}} = \frac{1}{2\pi}[\partial_\mu,
\partial_\nu] \, \omega_c
\]
The location of the (singular) Dirac worldsheet describes
an oriented two-dimensional manifold (in $\mathbb R^4$) at which
the transformation function $\omega_c$ has a singularity.
Since the string (in $\mathbb R^3$)
must be unobservable we arrive at the quantization of
the worldsheet pre-factor in~\eq{eq:f:law}
in terms of elementary magnetic charge $4 \pi/ e$.
In this manner the compactness of the internal group provides us with
the familiar identification of the electric charge,
\[
e = g \sin \theta_W\,.
\]

As an example, the Higgs field of a static Nambu monopole \cite{ref:Nambu}
is
\begin{equation}
\Phi = \eta f(r)
\left(
\begin{array}{c}
\cos \theta/2 \\
e^{i \varphi} \sin \theta/2
\end{array}
\right)
\lbl{eq:Phi:monopole}
\end{equation}
where $(r,\theta,\varphi)$ are spherical coordinates.
The singular structures (\ref{ds2}), (\ref{calG}) are
$\sigma^\phi_{\mu\nu} = 0$, $\Sigma^+_{\mu\nu} = 0$, and the string
\begin{equation}
\Sigma^3_{\mu\nu} = \delta(x_1)\delta(x_2) \theta(x_3)
[\delta_{\mu3}\delta_{\nu 4} - \delta_{\mu 4}\delta_{\nu 3}]
\lbl{eq:Nambu:sing}
\end{equation}
ends at the world line of the monopole, located at $r{=}0$.
The asymptotic behavior of the monopole's fields~\cite{ref:Nambu},
\begin{eqnarray}
g \vec W_\mu  & = & - \vec e_3 \times \partial_\mu
\vec e_3 - \vec e_3 \, \cos^2\theta_W \xi_\mu \,,
\\
g' Y_\mu  & = & \sin^2\theta_W \xi_\mu\,,
\end{eqnarray}
together with \eq{eq:Phi:monopole}
yields for the supercurrents ~\eq{J+}, \eq{J3} and \eq{Y} the asymptotic
behaviours
\begin{eqnarray}
J^+_\mu & = & 0 \\ g J^3_\mu & =
&g' {\mathcal Y}_\mu = \sin^2\theta_W\, \xi_\mu
\end{eqnarray}
so that asymptotically
\[
\begin{array}{rcl}
Z_\mu & = & 0 \\
A_\mu & = & \frac{\sin^2\theta_W}{e} \xi_\mu\,, \\
\xi_\mu & = & - i (\chi_1^\dagger \partial_\mu \chi_1 -
\partial_\mu \chi_1^\dagger \chi_1) =
(1 - \cos\theta) \partial_\mu \varphi
\end{array}
\]
where $\xi_\mu$ is the conventional field of the Dirac monopole.
The singularity structures show that the monopole
possess the non-Abelian charge $4 \pi/g$ while the magnetic hypercharge
(i.e. the charge with respect to the $Y$-field) is identically
zero, consistent with known results~\cite{ref:ana}.

In our variables, the gauge invariant 't~Hooft tensor \cite{thoof}
can be written as
\begin{equation}
{G}_{\mu\nu}
\equiv
\vec G_{\mu\nu} \cdot \vec e_3 - \frac{1}{g}
(\vec e_3 \cdot D_\mu \vec e_3 \times D_\nu \vec e_3)
= \partial_{[\mu,} J^3_{\nu]}
\lbl{eq:tHooft}
\end{equation}
where $D_\mu$ is the $SU_L(2)$ covariant derivative (\ref{covdev}).
The  't~Hooft tensor relates to the current $j^N_\mu$
that describes the world
trajectory $\mathcal C$ of the Nambu monopole,
\begin{equation}
\partial_\nu \tilde G_{\mu\nu} = \frac{4 \pi}{g} j^N_\mu
\ \equiv \  \frac{4 \pi}{g} \! \cdot \!\!
\int_{{\mathcal C}} {\mathrm{d}} y_\mu \, \delta^{(4)}(x - y)\,,
\end{equation}

The electroweak model also possesses various
string solutions~\cite{ref:ana} including
$Z$--vortices and $W$--vortices~\cite{ref:Z:string,ref:W:string} and
superconducting strings~\cite{ref:volkov}.
For example, the $Z$-string~\cite{ref:Z:string,ref:ana}
has a singularity only in the Abelian tensor (\ref{ds2}):
$\vec \Sigma_{\mu\nu} = 0$ and
\begin{equation}
\sigma_{\mu\nu}^\phi = \delta(x_1)\delta(x_2)
[\delta_{\mu 3}\delta_{\nu 4} - \delta_{\mu 4}\delta_{\nu 3}]\,,
\lbl{eq:Z:sing}
\end{equation}
while the W-string~\cite{ref:W:string,ref:ana} leads to
$\Sigma^3_{\mu\nu}=0$, $\sigma^\phi_{\mu\nu} = 0$ and
\begin{equation}
\Sigma^+_{\mu\nu} = e^{i \gamma} \delta(x_1)\delta(x_2)
[\delta_{\mu 3}\delta_{\nu 4} - \delta_{\mu 4}\delta_{\nu 3}]\,,
\end{equation}
where $\gamma$ is a phase.

Finally, we note that a representation of electroweak Lagrangian in
terms of gauge invariant variables has also been considered
in \cite{frohlich}. But the approach introduced there is quite different
from the present one. We also draw attention to
the strong coupling interpretation of electroweak Lagrangian proposed
in \cite{eddie}.

\section{Conformal Geometry}

We now propose the
Lagrangian (\ref{wsgi}) a generally covariant interpretation.
For this we analytically continue to Minkowski space with
signature $(-+++)$ and interpret $\rho^2$ in (\ref{wsgi}) as
a dilaton {\it i.e.} as the conformal scale
of a locally conformally flat metric tensor
\cite{faddeco},
\begin{equation}
\mathcal G_{\mu\nu} \ = \ \left( \frac{\rho}{\kappa}\right)^2 \eta_{\mu\nu}
\label{fadmet}
\end{equation}
Here $\eta_{\mu\nu}$ is the flat Minkowski metric. Since $\rho$ has
the dimensions of mass, we introduce
the {\it a priori} arbitrary mass parameter $\kappa$ to
ensure that the metric tensor has the correct dimensionality.

We accept the prescription in  \cite{hawk}, to
analytically continue the conformal scale (in the case
of asymptotically Euclidean manifolds \cite{hawk})
according to $\rho = 1 + \xi \to 1 -i \xi$, when
identifying the Minkowski signature metric
tensor.\footnotemark\footnotetext{We note that the prescription
has thus far been properly justified only in the case of
pure Einstein action.}
We then conclude that the Minkowski signature
Lagrangian (\ref{wsgi}) can be given the
following manifestly generally covariant interpretation,
\begin{equation}
\cL_\WS \ = \ \sqrt{- \mathcal G} \left\{ \frac{1}{16\pi G}(
R - 2\Lambda) + \cL_{M} \right\}
\label{egw1}
\end{equation}
with the matter Lagrangian $\cL_M$
\begin{equation}
\cL_{M}  =  - \frac{1}{4} \mathcal
G^{\mu\rho} \mathcal G^{\nu\sigma}
{\vec G}_{\mu\nu} {\vec G}_{\rho\sigma} - \frac{1}{4}
\mathcal G^{\mu\rho} \mathcal G^{\nu\sigma}
{F}_{\mu\nu} {F}_{\rho\sigma}
- \kappa^2 (g^2 + {g'}^2)
\mathcal G^{\mu\nu} Z_\mu Z_\nu -
\kappa^2 g^2
\mathcal G^{\mu\nu} W_\mu^+ W_\nu^-
\label{ewg2}
\end{equation}
We have here introduced parameters $G = 3/(8 \pi \kappa^2)$
and $\Lambda = (9\lambda)/(8\pi G)$
and for simplicity of notation the tensors $\vec G_{\mu\nu}$ and
$F_{\mu\nu}$ now contain (\ref{calG}) and (\ref{ds2}) respectively.

The result (\ref{egw1}), (\ref{ewg2}) re-interprets the
electroweak theory as a generally covariant gravity theory with
massive vector fields $Z$ and $ W^\pm$ and the (massless)
photon $A_\mu$.

Note that in (\ref{ewg2}) we have removed the (bare) Higgs mass
term that is present in (\ref{ew:lag}), as the Higgs mass term
is no longer neeeded in order for the theory to acquire
its desired physical properties.
In terms of the present variables the
correctly normalized masses for $Z$ and $W^\pm$ are
provided by the couplings $g$ and $g'$ and the
parameter $\kappa$, with no reference to the structure
of the Higgs potential and/or the mass of the Higgs.
We also note that the location of
structures such as vortex and monopole cores
where $\rho = 0$ can be interpreted in terms of
spacetime singularities.

We are now in a position to analyze the ground state structure of
the electroweak theory in the present variables. For this we first
note that in the ground state the massive vector
fields $Z_\mu$ and $W^{\pm}_\mu$ and the photon field $A_\mu$
all must vanish. Consequently the ground state is
determined by minimizing the gravitational contribution (\ref{egw1}).
This leads us to the de Sitter metric in its original form,
\[
ds^2 \ = \ \left( \frac{\rho^2}{\kappa^2} \right)
\eta_{\mu\nu} dx^\mu dx^\nu \ = \
\frac{ \eta_{\mu\nu} dx^\mu dx^\nu } { \left[
1 + \frac{4\pi}{9} G \Lambda \cdot x^2 \right]^2 }
\]
where the conformal scale is a solution to the following equation
of motion of ``$\lambda \phi^4$'' theory,
\[
- \Box \left(\frac{\rho}{\kappa}\right) - 8 \cdot \frac{4\pi}{9} G \Lambda
\left( \frac{\rho}{\kappa} \right)^3 = 0
\]
As a consequence in the present variables the ground state of
the electroweak theory is the de Sitter space.

If we recall that a four dimensional $\lambda \phi^4$
scalar field theory is trivial
\cite{froh} and adapt this result to
the present case, we conclude that the ``cosmological constant''
$\Lambda \to 0$. In this limit we recover the flat Minkowski space and the
ground state value of $\rho$ coincides with the parameter $\kappa$.
Thus, in this limit we arrive at the conventional symmetry
breaking picture of the original Weinberg-Salam model.

We also comment that if we do not follow the prescription in
\cite{hawk} the gravity Lagrangian acquires the
form
\[
\cL_{gravity} \ = \ \sqrt{- \mathcal G} \frac{1}{16\pi G}(-
R - 2\Lambda)
\]
This leads to a wrong sign in the Einstein equation in the presence of
matter fields. Now the ground state is anti-de Sitter space, and when
we employ stereographically projected coordinates we find
\[
ds^2 \ = \ \left( \frac{\rho^2}{\kappa^2} \right)
\eta_{\mu\nu} dx^\mu dx^\nu \ = \
\frac{ \eta_{\mu\nu} dx^\mu dx^\nu } { \left[
1 - \frac{4\pi}{9} G \Lambda \cdot x^2 \right]^2 }
\]
where the conformal scale emerges as a solution to the equation
of motion of the following ``$\lambda \phi^4$'' equation of motion,
\[
- \Box \left(\frac{\rho}{\kappa}\right) + 8 \cdot \frac{4\pi}{9} G \Lambda
\left( \frac{\rho}{\kappa} \right)^3 = 0
\]

We conclude this Section with the following comments:
The derivation of (\ref{egw1}), (\ref{ewg2}) employs the separation
between isospin and hypercharge. In parallel with spin-charge
separation in strongly correlated electron systems
\cite{anderson}-\cite{faddeco}
it becomes natural to interpret (\ref{egw1}),
(\ref{ewg2}) as a description of the electroweak theory in a
strongly coupled/strongly correlated (material) regime. From the point of
view of the original electroweak theory, the Lagrangian
(\ref{egw1}), (\ref{ewg2}) should now be interpreted in terms of
an effective Lagrangian which has been computed in the strongly
coupled/strongly correlated regime using the
covariant background field formalism.
The full effective Lagrangian accounts
for all quantum fluctuations in the fields of the original
tree-level electroweak Lagrangian (\ref{ew:lag}).
But since its explicit form is not available
beyond a few leading terms in a loop expansion, we have to resort to
an indirect analysis: By employing general arguments of gauge invariance
we expect that in terms of the original variables the full
effective Lagrangian is a functional of the background
fields and their background covariant derivatives. In the low momentum
infrared limit where we can ignore the higher order
derivative contributions, and since the full result is unknown to us,
we may for simplicity proceed by considering the infrared limit in
its lowest order.  This limit coincides with (\ref{egw1}), (\ref{ewg2}).
After all, the original classical Lagrangian {\it should} be an important
ingredient of the full quantum Lagrangian!

On the other hand, from the point of view of duality arguments
\cite{polch} it becomes attractive to view the gravity
Lagrangian (\ref{egw1}), (\ref{ewg2}) as a weak field and
short distance limit of a more complete gravity theory. For example,
the locally conformally flat form of the metric
tensor (\ref{fadmet}) could  be interpreted as the short
distance limit that emerges from a (renormalizable) higher
derivative gravity theory with a Lagrangian that contains the
following terms,
\[
{\mathcal L}_{W}^{} \ = \ \frac{\sqrt{-\mathcal G} }
{16 \pi G} \left( R -  2 \Lambda \right) \ + \
\sqrt{-\mathcal G} \cdot \gamma \hskip 0.9mm W^{2}_{\mu\nu\rho\sigma}
\]
Here $ W^{2}_{\mu\nu\rho\sigma}$ is the Weyl tensor.
In the short distance limit the one-loop $\beta$-function for
$\gamma$ sends this coupling to infinity \cite{stelle}.
This enforces asymptotically at short distances the condition
\[
W_{\mu\nu\rho\sigma} \ \sim \ 0
\]
which implies that locally, and in the absence of space-time
singularities, the short-distance metric tensor in this more complete
gravity theory assumes the conformally flat form (\ref{fadmet}).

\section{Summary}

In summary, we have shown that
in the Weinberg-Salam model the $SU_L(2) \times U_Y(1)$ gauge dependence
can be completely eliminated by a mere change of variables and without
any gauge fixing. As a consequence issues related to
Elitzur's theorem become obsolete.
The ensuing Lagrangian describes the electromagnetic
interactions of the gauge invariant and massive
$W$ and $Z$ bosons.
Furthermore, when we interpret the Higgs field as a dilaton
in a locally conformally flat spacetime, the electroweak
Lagrangian acquires a generally covariant
form and the vector bosons receive their correct masses
with no reference to any symmetry breaking by a Higgs potential.
Moreover, the ground state can be interpreted as the four
dimensional de Sitter space. However, this
interpretation assumes that we adopt the description of
\cite{hawk}. Otherwise, the ensuing Einstein equation
has a wrong sign for the matter coupling, and the gravity interaction
becomes repulsive with anti-de Sitter space as
the ground state of the theory.

We hope that our manifestly gauge invariant formulation of the
electroweak Lagrangian becomes valuable in properly interpreting
the structure of the electroweak transition which is soon to be
revealed at LHC.

\section*{Acknowledgements}

This work has been supported by a STINT Institutional grant IG2004-2 025.
M.N.Ch. was also supported by the grants NSh-679.2008.2, RFBR 08-02-00661-a
and RFBR\--DFG 06-02-04010 at early stage of this work.
The work by L.D.F. is also supported by grant RFBR 08-01-00638 and
program ``Mathematical problems of nonlinear dynamics'' of Russian
Academy of Science. The work by A.J.N is also supported
by a VR Grant 2006-3376 and by Project
Grant $\rm ANR~ NT05-1_{}42856$.
M.N.Ch. thanks M.S.Volkov for discussions. A.J.N.
thanks F. Wilc\-zek for a discussion.
Both M.N.Ch and A.N. thank S. Nicolis for a discussion.
M.N.Ch. is thankful to Laboratoire de Mathematiques et
Physique Theorique of Tours University
for hospitality. A.J.N. thanks the Aspen Center for Physics for hospitality.

\end{document}